\documentclass{elsart}

\usepackage{graphicx}

\usepackage{amssymb,amsmath}

\begin{document}

\begin{frontmatter}



\title{Threshold suppression of $\Lambda$ spin-orbit splitting}


\author[Gi]{Christoph M. Keil}
\ead{christoph.m.keil@theo.physik.uni-giessen.de}
\author[Gi]{Horst Lenske}
\ead{horst.lenske@theo.physik.uni-giessen.de}

\address[Gi]{Institut f\"ur Theoretische Physik, Heinrich-Buff-Ring 16, D-35392 Gie\ss en, Germany}

\begin{abstract}
New experimental data on medium to heavy single $\Lambda$ hypernuclei revealed a much
larger spin-orbit splitting than observed in older measurements of light hypernuclei.
Taking into account particle threshold effects and the density-dependence of in-medium
coupling constants the apparent suppression of spin-orbit strength 
in light hypernuclei as well as the spin-orbit structure observed
 in medium to heavy nuclei are explained in a unified manner within the
density dependent relativistic hadron field theory. It is concluded that the most valuable
information on the $\Lambda$ spin-orbit dynamics in finite nuclei has to be extracted from
medium to heavy mass nuclei.
\end{abstract}

\end{frontmatter}

\section{Introduction}
\label{sec:intro}
Guided by experiment, e.g. in $^{13}_{\Lambda}$C \cite{May:1981er} and 
$^{16}_{\Lambda}$O \cite{Bruckner:1978ix} it is a long standing opinion that 
the spin-orbit splitting of the $\Lambda$ single
particle levels in $\Lambda$ hypernuclei should be very small or almost zero. 
The most prominent theoretical explanation for
this effect has been given through the quark-spin substructure of the $\Lambda$ leading to
an additional $\Lambda$--$\omega$ tensor coupling that almost exactly cancels the 
more conventional scalar-vector spin-orbit 
force of the $\Lambda$ single particle states in finite systems 
\cite{Mares:1994xb,Chiapparini:1991ap,Tsushima:1998cu}.

A reanalysis of older $^{13}_\Lambda$C$^*$ \cite{Dalitz:1986xs} and $^{16}_\Lambda$O$^*$ 
\cite{Dalitz:1997nz} proton emitter emulsion data by Dalitz {\it et al.}, 
however, lead already to a remarkably
bigger splitting, i.e. $\approx$0.8 MeV for $^{13}_{\Lambda}$C and 1.56$\pm$0.12 MeV
for $^{16}_{\Lambda}$O though the statistics was not so good.
New experiments at KEK, measuring high resolution $\Lambda$ single-particle 
spectra for medium and heavy hypernuclei, show a spin-orbit splitting for the $\Lambda$ levels
in the range of 1--2 MeV \cite{Nagae99}. This is quite sizeable but still smaller
by a factor of about 2 than the splitting to be expected from the general quenching of
$\Lambda$--N interactions.

In this letter we will show that the reduced splitting
in medium and heavy nuclei can be attributed to the strong delocalization of the $\Lambda$ wave 
functions in the nucleus due to the shallow $\Lambda$ potential (sect.~\ref{sec:deloc}). 
In light nuclei with weakly bound $\Lambda$ states particle threshold effects are superimposed
on the general delocalization. The
very small splitting in light nuclei is due to these continuum threshold effects which lead
to a squeezing of the single particle levels close to threshold and thereby almost completely dilutes
generic spin-orbit potential effects in this mass region (sect.~\ref{sec:thresh}).
According to our investigations, the extremely small splitting of the spin-orbit
doublets in light $\Lambda$ hypernuclei as well as the reduced splitting of those in medium 
to heavy hypernuclei can be completely explained in the framework of the density dependent 
relativistic hadron (DDRH) field theory \cite{Keil:2000hk,Fuchs:1995as}.

\section{The DDRH field theory for $\Lambda$ hypernuclei}
\label{sec:DDRH}

\begin{figure}
\begin{center}
\includegraphics[width=60mm]{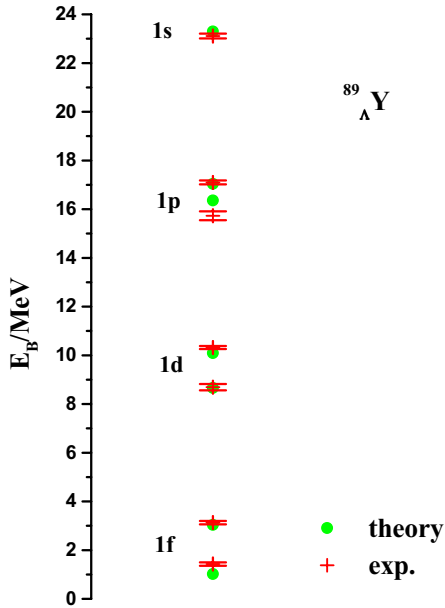}
\end{center}
\caption{Comparison of the measurements by Nagae et al. \cite{Nagae99} on $^{89}_{\Lambda}$Y with
DDRH calculations.}
\label{fig:expY}
\end{figure}

\begin{table}
\begin{center}
\begin{tabular}{l||r|r}
$^{51}_{\Lambda}$V&\multicolumn{2}{c}{Energies / MeV} \\
&Exp.&Theor. \\
\hline\hline
1s$_{1/2}$&-19.97$\pm$0.13&-19.89 \\
\hline
$\Delta$E(1p)&1.3$\pm$0.3&1.05 \\
E$_{centroid}$(1p)&--&11.5\\
$\Delta$E(1d)&2.0$\pm$0.2&1.81 \\
E$_{centroid}$(1d)&--&3.09
\end{tabular}
\end{center}
\caption{Comparison of the measurements by Nagae et al. \cite{Nagae99} on $^{51}_{\Lambda}$V with
DDRH calculations.}
\label{tab:expV}
\end{table}

The DDRH field theory \cite{Fuchs:1995as} is an effective density dependent
lagrangian field theory of nucleons and mesons based on microscopic nucleon--nucleon 
interactions. The extension of DDRH to the strangenes sector was described in \cite{Keil:2000hk},
where we refer to for theoretical details. Medium effects are included by mapping the ladder
sum of Dirac-Br\"uckner (DB) self-energies onto the vertices which are in our case functionals of
the baryonic field operators. As input for the vertices DB calculations for the Bonn A
potential have been used. Besides the standard kinetic and mass terms for baryons and mesons the DDRH 
lagrangian incorporates the interaction part:
\begin{eqnarray}
\mathcal{L}_{int} &=& \overline{\Psi}_F \hat{\Gamma}_\sigma(\overline{\Psi}_F, \Psi_F) \Psi_F \sigma
- \overline{\Psi}_F \hat{\Gamma}_\omega(\overline{\Psi}_F, \Psi_F) \gamma_\mu \Psi_F \omega^\mu \\
&&- \frac{1}{2}\overline{\Psi}_F \hat{\vec\Gamma}_\rho(\overline{\Psi}_F, \Psi_F) \gamma_\mu \Psi_F
\vec\rho^\mu
- e \overline{\Psi}_F \hat{Q} \gamma_\mu \Psi_F A^\mu \nonumber
\end{eqnarray}
Because of the functional dependence of the vertices on the field operators the baryonic
equations of motion contain additional 
{\it rearrangement self energies} accounting for static polarization effects of the nuclear medium
\cite{Keil:2000hk,Fuchs:1995as}.

In figure~\ref{fig:expY} and table~\ref{tab:expV} we compare our theoretical results with the 
new experimantal data
by H.~Hotchi and T.~Nagae \cite{Nagae99}. Our calculations have been performed with the
numerical parameter set derived from Dirac-Br\"uckner theory and free $\Lambda$-N scattering in 
\cite{Keil:2000hk}. The $\Lambda$ single particle 
spectra of $^{51}_\Lambda$V and $^{89}_\Lambda$Y are reproduced almost perfectly. This indicates 
that:
\begin{enumerate}
\item The dynamics of the $\Lambda$ is -- in medium to heavy nuclei -- almost completely governed
by the nuclear mean field which is in line with the experiment \cite{Nagae99} where shell structures
even for deeply bound $\Lambda$-states are clearly resolved.
\item In view of this not much room seems to be left for QCD related phenomena like
dissolving of the $\Lambda$ in the nuclear medium or 
partial deconfinement of the strange quark, as conjectured e.g. in \cite{Pirner:1979mb}.
\item A $\Lambda$--$\omega$ tensor coupling as it arises e.g. from the quark meson coupling
prescription and leads to an almost vanishing spin-orbit splitting (see e.g. 
\cite{Tsushima:1998cu}) is not supported by the analysis.
\end{enumerate}
This underlines the sensibility of a hadronic description of in-medium hyperons and their
interactions also around saturation density. It also stresses the
importance of $\Lambda$ hypernuclei in investigating the microscopic structure of
baryonic interactions.

\section{The spin-orbit potential and the delocalization of the $\Lambda$ wave function}
\label{sec:deloc}
As was indicated in \cite{Keil:2000hk} the reduced spin-orbit splitting in medium and
heavy $\Lambda$ hypernuclei can be attributed
to the delocalization of $\Lambda$ wave-functions. Since the spin-orbit energy is 
determined by the overlap integral of the single particle wave function and the spin-orbit
potential -- where the latter is well localized at the nuclear surface -- a strong delocalization
of the wave function leads to a sizeable reduction of the spin-orbit interaction energy.

\begin{table}
\begin{center}
\begin{tabular}{l||rr|rr|rr|rr}
&\multicolumn{2}{c}{$^{40}_{\Lambda}$Ca}&\multicolumn{2}{c}{$^{51}_{\Lambda}$V}
&\multicolumn{2}{c}{$^{89}_{\Lambda}$Y}&\multicolumn{2}{c}{$^{208}_{\Lambda}$Pb} \\
&$N_\Lambda$&$N_n$&$N_\Lambda$&$N_n$&$N_\Lambda$&$N_n$&$N_\Lambda$&$N_n$\\
\hline \hline
1p$_{3/2}$&0.51&0.71&0.59&0.75&0.68&0.81&0.79&0.88\\
1p$_{1/2}$&0.50&0.71&0.60&0.76&0.69&0.82&0.80&0.86\\
\hline
1d$_{5/2}$&0.23&0.47&0.33&0.54&0.47&0.64&0.66&0.80\\
1d$_{3/2}$&0.17&0.46&0.30&0.54&0.49&0.67&0.69&0.84\\
\hline
1f$_{7/2}$&--&--&--&--&0.27&0.45&0.52&0.72\\
1f$_{5/2}$&--&--&--&--&0.26&0.47&0.55&0.77\\
\end{tabular}
\end{center}
\caption{Localization coefficients $N_{\Lambda,n}(r_{rms})$ as defined in eq.~(\ref{eq:N}) for $\Lambda$
and neutron states. The reduced values for $\Lambda$ states indicate the increased
delocalization of $\Lambda$ wave functions.}
\label{tab:deloc}
\end{table}

A suitable measure for the reduced overlap of the $\Lambda$ wave function with the
nuclear interior is to define a localization coefficient by
\begin{equation}
\label{eq:N}
N_{\Lambda,n}(R)=N_o\int_0^R dr\;r^2\;\left|F_{\Lambda,n}(r)\right|^2
\end{equation}
describing the fraction of the probability density which is localized within the 
volume $\frac{4}{3}\pi R^3$.
Here, $F_{\Lambda,n}$ is the upper component of the respective Dirac wave function
and $N_o$ is chosen such that $N_{\Lambda,n}(R)\rightarrow 1$ for $R\rightarrow\infty$.
In table~\ref{tab:deloc} values $N_{\Lambda,n}(R)$ for $R=\sqrt{\left<r^2\right>}$ for $\Lambda$
and neutron orbitals in a number of nuclei are shown. The overlap of the $\Lambda$
wave functions with the bulk of the nuclear mass distribution is seen to be strongly
reduced compared to corresponding neutron states. The delocalization increases rapidly
with decreasing binding energy with a particular strong enhancement for $\Lambda$ states.
The results show clearly that a sizeable
part of the $\Lambda$ wave function lies ``outside'' the nucleus, a much larger fraction 
than in the neutron case. A comparison with experiment, 
see figure~\ref{fig:expY} and table~\ref{tab:expV}, shows that this mechanism is able to
explain the fine structure of $\Lambda$ single particle spectra.

\section{Continuum threshold effects on the spin-orbit splitting}
\label{sec:thresh}
Experimentally observed single particle spectra in low mass $\Lambda$ hypernuclei show a much 
more drastic reduction of the spin-orbit splitting. This led to the conjecture
that the spin-orbit splitting in hypernuclei should be in general very small and could be 
reasonably explained by assuming that the spin of the $\Lambda$ is carried solely by the
strange quark.

The level spacing in light nuclei is, however, strongly affected by the continuum threshold
which is in these small systems -- especially for the rather weakly bound $\Lambda$s -- quasi
omnipresent. As is known from neutron rich nuclei, e.g. \cite{rm2000,Le,19c,8b},
the single particle level
spacing becomes compressed for states approaching the threshold.

\begin{figure}[htb]
\begin{center}
\includegraphics[width=78mm]{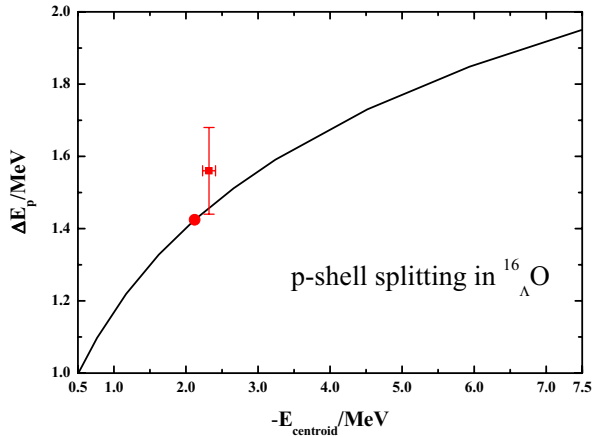}
\end{center}
\caption{The figure shows the strong decrease of the spin orbit splitting for the 1p doublet
in $^{16}_\Lambda$O when the doublet's centroid is pushed towards the continuum threshold.
The dot marks the actual theoretical value for $^{16}_\Lambda$O, the data point is taken 
from \cite{Dalitz:1997nz}.}
\label{fig:thresh}
\end{figure}

The threshold compression effect is visible already in the Ca region but is especially important
around and below $^{16}_\Lambda$O. The typical feature in these nuclei is that the centroid
energy of a spin-orbit doublet is rather close to threshold, mainly because of the
shallowness of the binding potential. Under such conditions the spectrum of valence levels
becomes compressed inducing an apparent reduction of spin-orbit strength because the less
bound j=$\ell$-1/2 member of a spin-orbit doublet tries to avoid the crossing in the unbound 
region. This ``hindrance'' is related to the general phenomenon of ``avoided level crossing''
in quantal systems
because passing over into the continuum means that for example a $p_{1/2}$-level would
have to cross the region of low energy s-wave continuum levels before re-appearing as a
single particle resonance.

We illustrate these special effects for the 1p spin-orbit doublet in $^{16}_{\Lambda}$O.
Approaching the continuum threshold is enforced by modifying artificially the $\Lambda$ 
potential depth by variation of the $\Lambda$--$\omega$ 
coupling constant. Figure~\ref{fig:thresh} shows the dependence of the splitting strength of
the 1p doublet on the centroid energy $E_{\text{centroid}}=\frac{E_{p3/2}+E_{p1/2}}{2}$. 
Even strongly bound states (E$_B<$-6MeV) are still considerably affected by the threshold. The 
effect only saturates if a state of another parity from the continuum is lowered into the 
bound region, in this example a $1d_{5/2}$ orbit. The same behavior is also found in the 
1p-shell doublet in $^{13}_\Lambda$C, often referred to as the ``standard hypernucleus''.

The continuum threshold effect reflects the peculiar dependence of
matrix elements on wave function overlaps for weakly bound states.
In such cases - as observed in recent studies of pure isospin
dripline nuclei - generic properties of interactions are masked
behind wave function effects. This has important consequences for
extracting the desired information on the underlying interaction
from data: The threshold effect leads to local variations in
energy splittings not following the systematics derived from well
bound systems. Obviously, this will strongly inhibit a model
independent derivation of $\Lambda$ interaction strengths from
measurements, especially in light nuclei. Around and beyond
A$\sim$40, however, the threshold effect rapidly vanishes. Since
the new KEK data \cite{Nagae99} for $^{51}V$ and $^{89}Y$ were obtained
in mass region well outside the threshold regime they can be
expected to provide suitable information on the generic $\Lambda$
spin-orbit strength, indicating a larger value than assumed
before.

\section{Conclusion}
The larger than expected spin-orbit splitting 
observed in recent KEK experiments on medium mass nuclei are well
reproduced by relativistic DDRH mean-field calculations. Universal
parameter sets, derived previously from Dirac-Brueckner theory,
were used. The results indicate that the medium dependence of
hyperon-coupling constants is essential for the good description.
The calculations explain consistently also the apparent
suppression of spin-orbit strength for weakly bound $\Lambda$
orbitals in light nuclei by the compression of spectra close to
the continuum threshold due to avoided level crossing. It is
worthwhile to emphasize strongly the importance of high-resolution
data, like those of ref.~\cite{Nagae99}, for hypernuclear structure theory and
the understanding of hyperon interactions.

\section{Acknowledgements}
This work was supported by DFG grant Le439/4-3 and the European Graduate School ``Complex
Systems of Hadrons and Nuclei, Gie\ss en--Copenhagen''.

\end{document}